\documentclass[twocolumn]{article}
\usepackage{times}
\usepackage{amssymb}
\usepackage{epsfig}
\usepackage{array}
\usepackage{amsmath}
\usepackage{endnotes}

\newcommand{\Lya}{Ly$\alpha$}

\title{The Lyman-$\alpha$ glow of gas falling into the dark matter halo of a $z=3$ galaxy}
\date{}
\author{M.~Weidinger$^{1,2}$, P.~M{\o}ller$^1$, J.~P.~U.~Fynbo$^{2,3}$\\
{\small $^1$ European Southern Observatory, Karl-Schwarzschild-Stra{\ss}e 2,
D-85748 Garching bei M\"unchen, Germany}\\
{\small $^2$ Institute of Physics and Astronomy, University of Aarhus, Ny
Munkegade, DK-8000 {\AA}rhus C, Denmark}\\  
{\small $^3$ Astronomical Observatory, University of Copenhagen, Juliane
  Maries Vej 30, DK-2100 Copenhagen {\O}, Denmark} 
}

\begin{document}
\maketitle

{\bf \noindent Quasars are the visible signatures of super-massive
black holes in the centres of distant galaxies. 
It has been suggested\endnote{e.g.~Carlberg, R.G. Quasar
evolution via galaxy mergers. {\it Astrophys. J.} {\bf 350}, 505-511
(1990)} that quasars are formed during ``major merger events'' when two
massive galaxies collide and merge, leading to the prediction
that quasars should be found in the centres of the regions of largest
overdensity in the early Universe. In dark matter (DM)-dominated
models of the early Universe, massive DM halos are predicted to attract
the surrounding gas, which falls towards its centre. The neutral gas is
not detectable in emission by itself, but gas falling into the ionizing
cone of such a quasar will glow in the Lyman-$\alpha$ line of hydrogen,
effectively imaging the DM halo\endnote{Haiman, 
Z. \& Rees, M. Extended {\Lya} Emission around Young Quasars: A
Constraint on Galaxy Formation. {\it Astrophys. J.} {\bf 556}, 87-92
(2001)}. Here we present a {\Lya} image of a DM halo at redshift
3, along with a two-dimensional spectrum of the gaseous halo.
Our observations are best understood in the context of the
standard model for DM halos\endnote{Navarro, J.F.,
Frenk, C.S., \& White, S.D.M. A Universal Density Profile from
Hierarchical Clustering. {\it Astrophys. J.} {\bf 490}, 493-508
(1997)}; we infer a mass of $(2-7)\times 10^{12}$ solar masses
($M_\odot$) for the halo. }

Using radiative transfer calculations, Haiman \& Rees$^{\rm ref 2}$
predicted that gas falling into a DM halo between redshifts $3$
and $8$ that was harbouring a quasar should be detectable in \Lya-emission at
flux levels accessible to present day telescopes, owing to the
reprocessing of quasar ultraviolet photons. Recently, Barkana \&
Loeb\endnote{Barkana, R. \& Loeb, A. Spectral signature of cosmological
infall of gas around the first quasars. {\it Nature} {\bf 421}, 341-343
(2003)} found absorption features in quasar spectra, which they
interpreted as a signature of neutral hydrogen (H{\sc i}) falling into
the DM halos surrounding two quasars. Such absorption features can be
used to study the gas in one dimension (1D) along the line of sight,
whereas detection of extended \Lya-emission can be used in a 2D study,
providing more
constraints on the interplay between gas, quasar radiation and DM halo.

\begin{figure}[p]
\centering\epsfig{file=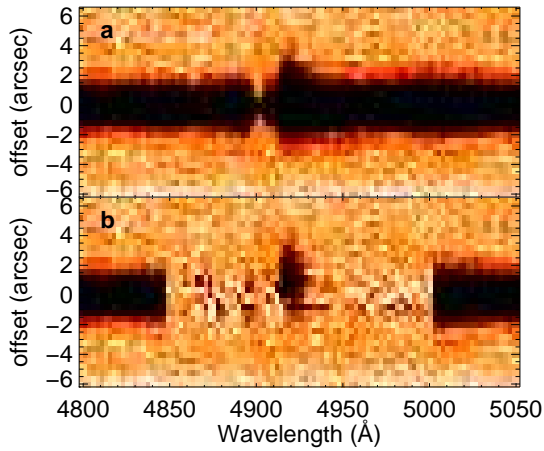,width=8cm}
\caption{{\small Two-dimensional spectrum of Q1205-30 and the extended
  {\Lya} emission. The
  wavelength increases along the abscissa, the position on the sky
  changes along the ordinate. The spectroscopy was performed with a
  slit  position angle of $7.9^\circ$ east of north (see
  Fig.~\ref{fig:field}a). {\bf a}, The 2D quasar spectrum. The
  extended {\Lya} emission is faintly visible at $4,920$ {\AA}. {\bf b},
  The quasar spectrum has been subtracted between the wavelengths $4,850
  - 5,000$ {\AA}, clearly revealing the underlying extended
  emission.}}\label{fig:fig2D}
\end{figure}

\begin{figure}[p]
\centering\epsfig{file=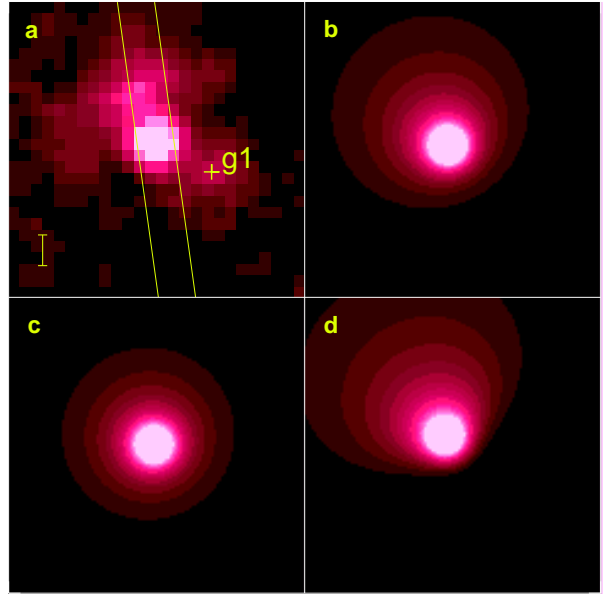,width=8cm}
\caption{{\small Narrow-band image of the extended {\Lya} emission compared to
  models. {\bf a}, Image of the $10\times 10$ arcsec$^2$ region around
  Q1205-30 seen in \Lya. The quasar has been subtracted from this image
  to reveal the underlying extended emission. A $1.2$-arcsec-wide slit
  is overplotted at a position angle of $7.9^\circ$ east of north, and the
  position of an unrelated foreground galaxy$^{\rm 5}$ is marked
  g1. The vertical bar on the lower left is $1$ arcsec high. {\bf b --
  d}, Calculated $10\times 10$ arcsec$^2$ surface brightness maps for
  an opening angle of $\Psi = 110^\circ$ and three inclination
  angles. The best fitting inclination angle ($\theta = 27^\circ$) is
  shown in {\bf b}, two other inclination angles are shown in {\bf c}
  ($\theta=10^\circ$, excluded at $3.5\sigma$) and {\bf d}
  ($\theta = 50^\circ$, excluded at $3\sigma$) for
  comparison. Given an opening angle, the best-fit inclination angle is
  determined in the following way. An average surface brightness
  profile, parallel and perpendicular to the apparent major axis of the
  extended emission, is fitted to the same average calculated from the
  models. For each opening angle $\Psi$ the best-fitting inclination
  angle is found.}}\label{fig:field}
\end{figure}

In a deep {\Lya} narrow-band image we detected asymmetric extended
emission North-East of the $z=3$ radio-quiet quasar Q1205-30, making
this quasar a prime candidate for a study of its DM halo. The extended
{\Lya} emission was first thought to be related to a foreground
absorber\endnote{Fynbo, J.P.U., Thomsen, B., \& M{\o}ller, P. {\Lya}
emission from a Lyman limit absorber at z=3.036. {\it
Astron. Astrophys.} {\bf 353}, 457-464 (2000)}, but deep follow-up
spectroscopy obtained with the FORS1 instrument on the ESO Very
Large Telescope allowed us to measure precise redshifts of quasar,
absorber, and extended emission, clearly linking the extended emission
to the quasar, not the absorber. The 2D spectrum is presented in
Fig.~\ref{fig:fig2D} (details of the data reduction will be presented
elsewhere). In the 2D spectrum the wavelength increases along the
abscissa, and the position on the sky changes along the ordinate. In
order to reveal the underlying extended emission we have optimally
extracted the 2D quasar spectrum by fitting and removing the quasar
spectral point-spread function\endnote{M{\o}ller, P. Spectral PSF
Subtraction I: The SPSF Look-Up-Table Method. {\it The ESO Messenger}
{\bf 99}, 31-33 (2000)}. Spectroscopic detection of kinematically
resolved extended \Lya-emission is not uncommon\endnote{M{\o}ller, P.,
Warren, S.J, Fall, S.M., Jakobsen, P., \& Fynbo, J.U. SPSF Subtraction
II: The Extended {\Lya} Emission of a Radio Quiet QSO. {\it The ESO
Messenger} {\bf 99}, 33-35 (2000)}$^{,}$\endnote{Bunker A., Smith, J.,
Spinrad, H., Stern, D., \& Warren., S.J. Illuminating Protogalaxies?
The Discovery of Extended Lyman-$\alpha$ Emission around a QSO at
$z=4.5$. {\it Astrophys. Sp. Sci.} {\bf 284}, 357-360 (2003)}, but it
does not provide sufficient information to distinguish between
scenarios of -- for example -- nearly edge-on disks and gaseous
halos. For this, morphological information is needed. In
Fig.~\ref{fig:field}a we present a $10\times 10$ arcsec$^2$
narrow-band image of the extended emission.

\begin{figure}[p]
\centering\epsfig{file=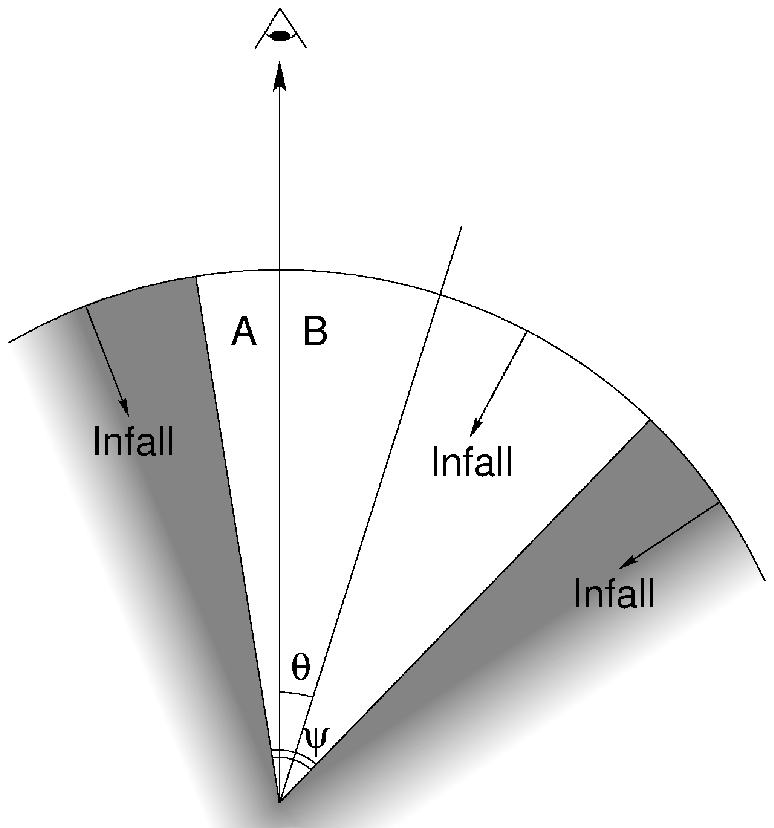,width=8cm}
\epsfig{file=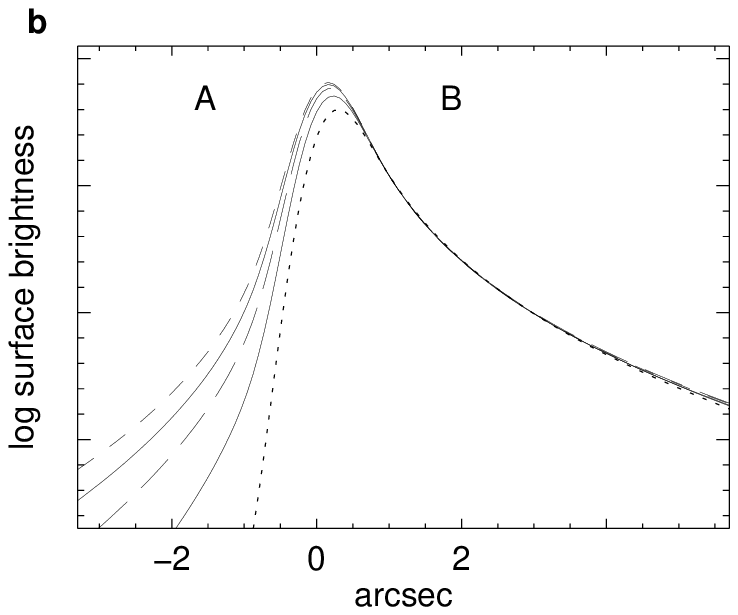,width=8cm}
\caption{{\small Schematic presentation of the model. {\bf a}, A
  quasar is located in the centre of a DM halo into which neutral
  hydrogen is falling. The ionizing photons from the quasar are emitted
  in a cone of opening angle $\Psi$, and they cause the infalling H{\sc
  i} to glow in \Lya. The observer views the system inclined at an
  angle $\theta$. The surface brightness at each point is calculated by
  integration of the volume emission along the line of sight$^{\rm ref 9}$:
  $\Sigma_{{\rm Ly}\alpha} = \frac{\int E_{{\rm Ly}\alpha}\dot{n}_{{\rm
  Ly}\alpha}(r) dl}{4\pi D_L^2} \frac{D_A^2}{d\Omega}$, where $E_{{\rm
  Ly}\alpha}=10.4$ eV is the energy of a {\Lya} photon, $\dot{n}_{{\rm
  Ly}\alpha}(r)$ is the production rate of {\Lya} photons at a distance
  $r$ from the quasar, $D_L$ is the quasar luminosity distance, and
  $D_A$ is the angular distance, such that $\frac{d\Omega}{D_A^2}$ is
  the conversion from cm$^2$ to arcsec$^2$. The {\Lya} production rate
  is $\dot{n}_{{\rm Ly}\alpha}(r) = \eta_{\rm thin} n_{\rm HI}(r)
  \Gamma(r)$, and $\eta_{\rm thin} =0.42$ is the probability for an
  ionizing photon to result in a {\Lya} photon in the optically thin
  case, $n_{\rm HI}(r)$ is the volume density of neutral hydrogen
  atoms, and $\Gamma(r)$ is the ionization rate of hydrogen atoms.
  {\bf b}, A schematic illustration of the predicted surface
  brightness profile for various opening angles $\Psi = 90^\circ$
  (dotted), $110^\circ$ (solid), $130^\circ$ (long-dashed), $150^\circ$
  (solid), and $170^\circ$ (short-dashed), and a fixed inclination
  angle $\theta = 40^\circ$. The profiles
  are presented as they would appear when observed with a seeing of
  $0.7$ arcsec through a slit aligned with the line between the
  sightline $l$ and the symmetry axis of the cone. Negative angular
  distances correspond to emission seen from side A, and positive
  angular distances correspond to emission seen from side
  B.}}\label{fig:model}
\end{figure}

A quasar emits its radiation in an ionizing cone with an opening angle
of $\Psi$, while we observe the whole system under an inclination angle
of $\theta$ (see Fig.~\ref{fig:model}). The surface brightness at each
point in the vicinity of the quasar is calculated by integration of the
volume emission along the line of sight\endnote{Gould, A. \& Weinberg,
D.H. Imaging the Forest of Lyman Limit Systems. {\it Astrophys. J.}
{\bf 468}, 462-468 (1996)}. We take the gas density profile to be a
power-law, $n_{\rm HI}(r) = n_{\rm HI, 1} \left(r/1\text{ kpc}
\right)^{-\alpha}$, with a slope $\alpha$ and a neutral hydrogen
density at a distance of $1$ kpc $n_{\rm HI, 1}$, because this appears in
numerical simulations of DM halos to be a good approximation at small
radii\endnote{Barkana, R. A model for infall around virialized
haloes. {\it Mon. Not. R. Astron. Soc.} {\bf 347}, 59-66 (2004)}. If we
assume a mass of the DM halo, we may calculate the observed infall
velocity in the model. At a given projected distance from the quasar
the observed velocity is the average projected velocity weighted with
the emissivity of the gas along the line of sight.

Applying this model to Q1205-30, our calculation shows that the
Ly$\alpha$ halo should be detectable with 8-m class telescopes up to
several arcsec from the quasar, assuming a modest amount of neutral
hydrogen infall. Considering -- for now -- the case where only one cone is
visible, the extended emission may be symmetric (if $\theta\approx 0$,
that is, the cone is seen close to end-on) or highly asymmetric (if
$\theta\approx \Psi/2$). By fitting the calculated surface brightness
maps to the observed narrow-band image we establish a relation between
the opening angle and the best-fitting inclination angle. Thus, we are
left with the opening angle, $\Psi$, the slope, $\alpha$, and the
neutral hydrogen density at $1$ kpc, $n_{{\rm HI}, 1}$, as the only
free parameters of the model. In Fig.~\ref{fig:field}, we compare our
observed narrow-band image to calculated {\Lya} surface brightness maps
with an opening angle $\Psi=110^\circ$ and three different inclination
angles: the best-fit and two others. For the expected large opening
angles ($\Psi \approx 90^\circ - 120^\circ$)\endnote{Lawrence, A. The
relative frequency of broad-lined and narrow-lined active galactic
nuclei - Implications for unified schemes. {\it
Mon. Not. R. Astron. Soc.} {\bf 252}, 586-592
(1991)}$^{,}$\endnote{Elvis, M. A Structure for Quasars. {\it
Astrophys. J.} {\bf 545}, 63-76 (2000)}, varying the opening angle
primarily affects the shape of the surface brightness profile seen on
the side where the emission is weakest (in Fig.~\ref{fig:model}, that
would be on side A of the sightline), and only has a minor effect on
the shape of the main emission profile (on side B). The gas density
scale, $n_{{\rm HI}, 1}$, determines the normalization of the surface
brightness profile, but not the shape, which is set by the slope
$\alpha$. The results are not dependent on the choice of a power-law
gas density profile. An exponential gas density profile provides an
equally good fit to the data.

\begin{figure}[p]
\centering\epsfig{file=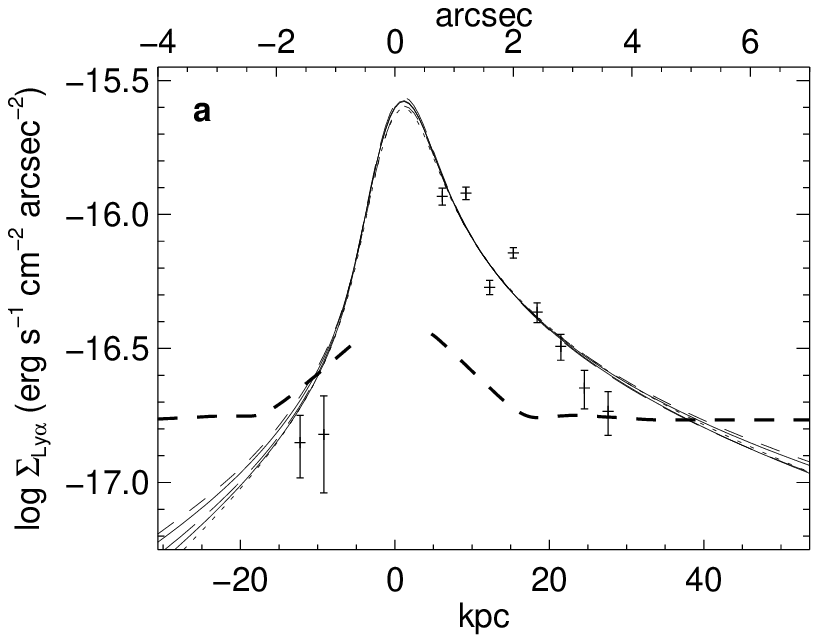,width=8cm}
\epsfig{file=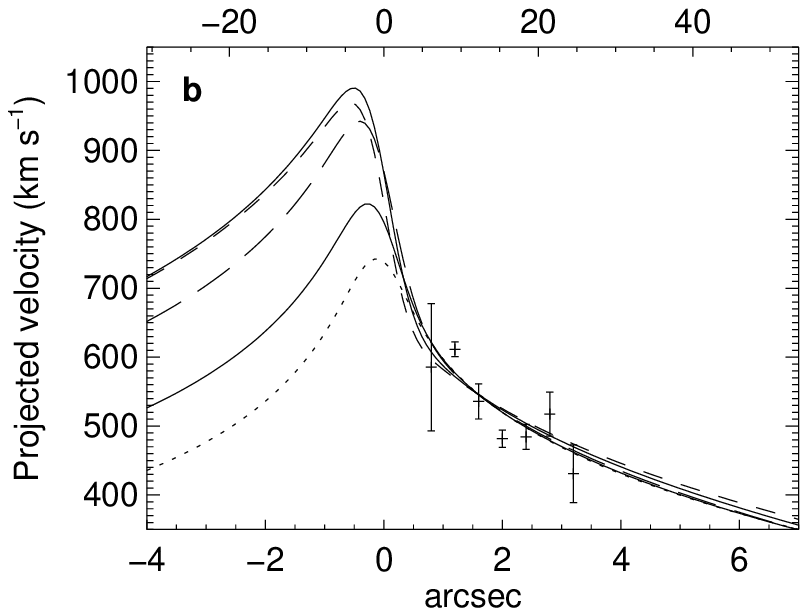,width=8cm}
\caption{{\small Calculated surface brightness profile and projected
  velocity. {\bf a}, Comparison of the calculated and observed surface
  brightness profile. Thin lines are the calculated profile for various
  opening angles, and data points are our observations of the extended
  \Lya-emission around Q1205-30. The calculations were performed for
  opening angles and inclination angles of $\Psi = 90^\circ$,
  $\theta = 22.5^\circ$ (dotted), $\Psi = 110^\circ$ , $\theta =
  27.2^\circ$ (solid), $\Psi = 130^\circ$, $\theta = 31.8^\circ$
  (long-dashed), $\Psi = 150^\circ$, $\theta = 36.5^\circ$ (solid), and
  $\Psi = 170^\circ$, $\theta = 41.1^\circ$ (short-dashed). The thick
  dashed line shows our $5\sigma$ detection limit. {\bf b}, Comparison
  of calculated and observed projected velocity relative to the quasar
  redshift. Lines are the calculated best-fit velocity profiles for
  opening angles and inclination angles of $\Psi = 90^\circ$,
  $\theta = 22.5^\circ$ (dotted), $\Psi = 110^\circ$, $\theta =
  27.2^\circ$ (solid), $\Psi = 130^\circ$, $\theta = 31.8^\circ$
  (long-dashed), $\Psi = 150^\circ$, $\theta = 36.5^\circ$ (solid), and
  $\Psi = 170^\circ$, $\theta = 41.1^\circ$ (short-dashed). Three of
  the measurements in panel {\bf a} were too faint to allow a secure
  velocity determination. In both panels the angular distance was
  converted to physical distance assuming a flat $\Omega_\Lambda = 0.7$
  Universe with a Hubble constant $H_0 = 70\text{ km
  s}^{-1}\text{Mpc}^{-1}$. Error bars are $1\sigma$.} }\label{fig:SBvel}
\end{figure}

The extended emission is spatially resolved, so we can measure its
surface brightness profile and velocity profile (see
Fig.~\ref{fig:SBvel}). The surface brightness profile is measured by
integrating the flux from $4900$ {\AA} to $4947$ {\AA} in each spatial
bin. The velocity in each spatial bin is measured by fitting a gaussian
to the line profile. The infall velocity is calculated relative to the
redshift $z=3.041$ of the quasar.

We find that the model describes the observations well. The opening
angle is the only parameter which remains unconstrained, so we plot the
surface brightness profile for a range of opening angles
(Fig.~\ref{fig:SBvel}a). The best-fit values for the neutral hydrogen
density at $1$ kpc vary less than a factor two between $\Psi=90^\circ$
and $170^\circ$, while the slope varies between $0.02$ and $0.09$. The
velocity profile of the extended emission (Fig.~\ref{fig:SBvel}b)
arises as a projection effect, and fitting velocity curves of canonical
DM halo profiles$^{\rm ref 3}$ to the observed velocities, we infer a
virial mass of $(2-7)\times 10^{12} M_\odot$ for the DM halo. An
identical mass estimate is obtained when using an exponential gas
density profile. The H{\sc i} density from our best fitting model
allows us to calculate an accretion rate of $\sim 0.1 M_\odot$yr$^{-1}$
in neutral hydrogen. The total accretion rate will be higher as the gas
is highly ionized.

Where studies based only on spectra are unable to provide evidence to
distinguish between several different scenarios, the combination of
deep imaging and deep spectroscopy allows us to rule out alternative
explanations of the extended emission. Jets are generally believed to
be present in radio-quiet quasars and are predicted to extend out to
$\sim 0.1$ kpc\endnote{Blundell, K.M., Beasley, A.J., \& Bicknell,
G.V. A Relativistic jet in the radio-quiet quasar PG 1407+263. {\it
Astrophys. J.} {\bf 591}, L103-106 (2003)}, where the extended
{\Lya} emission around Q1205-30 extends out to $\sim 30$ kpc. Outflowing
galactic winds are believed to be triggered by supernovae going off
inside the galaxy, and are therefore expected to be metal-enriched. Our
detection limit of $4\times 10^{-18}$ erg
s$^{-1}$ cm$^{-2}$ arcsec$^{-2}$ ($3\sigma$) ensures that we are able
to detect the C{\sc iv} and He{\sc ii} lines typical in extended
emission around radio-loud quasars\endnote{Heckman, T.M., Lehnert,
M.D., Miley, G.K, \& van Breugel, W. Spectroscopy of spatially extended
material around high-redshift radio-loud quasars. {\it Astrophys. J.}
{\bf 381}, 373-385 (1991)}. We did not detect any of these lines, so the
gas appears not to have been enriched by supernovae.

{\small Recieved 24 March; accepted 28 June 2004.}

\theendnotes
\bigskip

\noindent{\bf Correspondence} and requests for materials should be
addressed to M.W. (mweiding@eso.org).

\noindent{\bf Acknowledgements} Based on observations made with ESO
Telescopes at the Paranal Observatory.

\noindent{\bf Competing interests statement.} The authors declare that
they have no competing financial interests.

\end{document}